# Complete description of re-entrant phase behaviour in a charge variable colloidal model system

Patrick Wette[1], Ina Klassen[1], Dirk Holland- Moritz[1], Dieter M. Herlach[1], Hans Joachim Schöpe[2], Nina Lorenz[2], Holger Reiber[2], Thomas Palberg[2] and Stephan V. Roth[3]

[1]*Institut für Materialphysik im Weltraum, Deutsches Zentrum für Luft- und Raumfahrt (DLR), 51170 Köln, Germany*

[2]*Institut für Physik, Johannes Gutenberg Universität Mainz, Staudinger Weg 7, 55128 Mainz, Germany*

[3]*HASYLAB, DESY, Notkestraße 85, 22603 Hamburg, Germany*

**Abstract**

In titration experiments with *NaOH* we have determined the full phase diagram of charged colloidal spheres in dependence on the particle density $n$, the particle effective charge $Z_{eff}$ and the concentration of screening electrolyte $c$ using microscopy, light and Ultra Small Angle X-Ray Scattering (USAXS). For sufficiently large $n$ the system crystallizes upon increasing $Z_{eff}$ at constant $c$ and melts upon increasing $c$ at only slightly altered $Z_{eff}$. In contrast to earlier work equilibrium phase boundaries are consistent with a universal melting line prediction from computer simulation, if the elasticity effective charge is used. This charge accounts for both counter-ion condensation and many body effects.

The phase behaviour of charged colloidal spheres in aqueous suspension has attracted a lot of interest for their great potential use as model systems for the (classical) properties of metals[1]. Such Yukawa-like systems interact via a screened Coulomb potential can be varied over a wide range between the theoretical limits of the one component plasma and the hard sphere system[2]. The experimentally tuneable parameters are the particle number density $n$, the concentration of screening electrolyte $c$, and the effective charge $Z_{eff}$. Early measurements focused on variations of $n$ and $c$, reporting face centred cubic (fcc) structures at large $n$ and later also body centred cubic (bcc) structures at low $n$ and low salt conditions [3,4,5,6]. A complete phase diagram of the behaviour at moderate to large $c$, covering also large $n$ was measured by Sirota et al.[7] using Ultra Small X-Ray Scattering (USAXS). There also a glass phase was reported for the first time for charged sphere colloids. In all these experiments the bare particle charge Z stemming from particle synthesis was held constant. The influence of a variable charge is more subtle and was addressed much less frequently. There are three main effects to be considered: Counter-ion condensation, self-screening and many body effects appearing as macro-ion shielding. A condensation of counterions at the particle surface occurs for large surface charge densities. Bucci et al.[9] were the first to demonstrate the implications for structure formation in an experimental study on in mixed micelles. They found that consistent with theory[8] the effective charge $Z_{eff}$ would saturate at a certain value while Z was further increased[9] and no further gain of structure could be obtained. The effect was then elaborated by both theory[10] and experiment[11,12,13,14] and its interpretation as counter-ions condensation is generally accepted.[15] The corresponding number of free counter-ions is accessible from conductometry[16,17]. The phase behaviour of Yukawa systems was first investigated with extensive computer simulations by Robbins, Kremer and Grest (RKG), where a universal melting line for charged colloids was observed.[18] Later the study by Meijer and Frenkel (MF)[19] leads to a melting line predicting a lower crystal stability compared to RKG whereas simulations by Hamaguchi et al.[20] gave results located between the ones of RKG and MF. Early studies of crystallization in charge variable systems[21] reported a good agreement with the RKG predictions. Later studies at larger bare charge, however showed significant deviations towards a decreased crystal stability[5,22,23] when the charge from conductivity measurements was used. Moreover, upon charging up the particles by adding *NaOH* the system first froze then re-melted. This could only partly explained as being due to an increased screening by the increase in the number of free counter-ions upon charging the particles (self screening). The authors suggested like charge attraction and many body effects as additional alternative explanations.

Recent experiments clearly favour the latter possibility. For a long ranged interacting 2D charged sphere fluid a significant reduction of repulsions was observed and attributed to macro-ion shielding.[24] The cage of nearest neighbours leads to a cut off in the repulsion, as soon as the interaction range exceeds the nearest neighbour distance. This clear indication of many-body forces has since attracted considerable interest and drastic consequences for the phase behaviour were predicted.[25,26] A comparison of the different experimentally and computationally accessible charge numbers was given by Shapran et al.[27] The effective charge under shielding conditions is given by measuring the elasticity of polycrystalline solids. In fact, Wette and Schöpe had shown that a consistent description of the phase transition for particles of constant bare charge $Z$ is obtained, if this charge is used instead of the conductivity effective charge.[14] Later Shapran et al found from elasticity measurements along the phase boundary indications of a transition from a condensation only to a condensation plus shielding regime upon decreasing $n$ and $c$.[28] At present it is a completely open issue, whether this concept may also be applied to charge variable particles.

The present paper goes beyond previous studies in several ways: We perform a combined microscopy, light scattering and USAXS study on the phase behaviour of silica particles in dependence on $n$ and the amount of added *NaOH*. The combination of different methods studying the systems structure considerably enhances the range of particle concentrations. We combine different ways to tune the particle interaction in one single system allowing a more reliable consistency check of the mean field description of charged colloidal crystal properties using the interaction potential determined by elasticity measurements.[14] We confirm the re-entrant shape of the phase diagram from earlier studies[29] with much improved statistics and a qualitative extension to the regime of melting by the added electrolyte. In addition we measure the elasticity effective charge of the polycrystalline solids along the phase boundaries and use these to compare to the theoretical expectations of RKG. We find a good agreement with the theoretical predictions *without* evoking like charge attraction arguments, when many body effects are accounted for by using this elasticity effective charge.

Silica particles of diameter *2a = 84nm* were synthesized by Stöber synthesis and conditioned using an automated version of a closed-circuit conditioning set-up reported earlier [30]. The peristaltically driven tubing circuit connects an ion exchange column, a reservoir under Argon atmosphere to add water, particles and *NaOH,* a conductivity cell to control the state of deionization and/or charging, the scattering cell and a cell to measure elasticity by torsional resonance spectroscopy for transparent samples of $n<60\mu m^{-3}$. It thus ensures identical interaction conditions for all measurements including the determination of the conductivity effective charge and the elasticity effective charge.[14] Fortunately, the polycrystalline solids are isotropic for length scales corresponding to the wave length of the excited elastic waves and as we determine the orientationally averaged elasticity we may interpret our measured shear moduli in terms of an effective pair potential.[28] The cell is sealed from the preparation circuit by electromagnetic valves upon start of the structure measurements and the sample readily re-crystallizes upon stop of shear. The phase behaviour was determined at low $n$ using standard static light scattering (open symbols in Fig. 2). At larger $n$ we performed USAXS measurements at the BW4-beamline at HASYLAB Hamburg at a wavelength of $\lambda=0.138nm$ and a sample to detector distance of $13.3m$ leading to a wide scattering vector range of $0.008 < q < 0.280nm^{-1}$ (closed symbols).

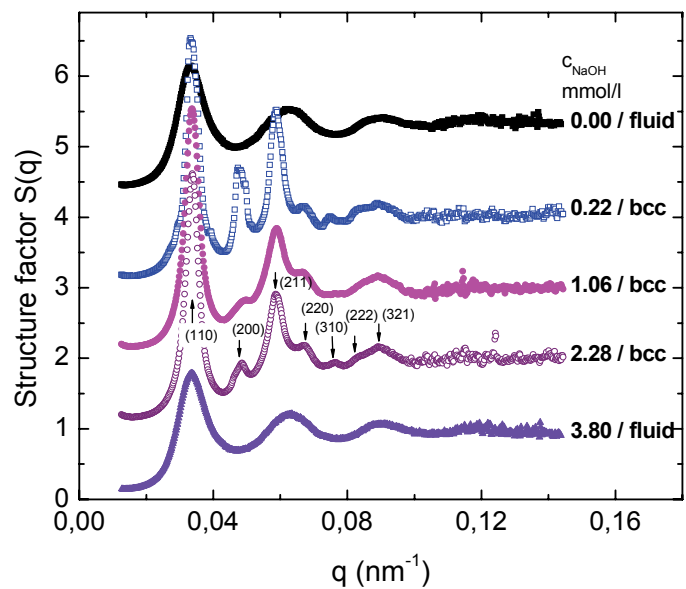

Fig. 1: (Color online) *USAXS static structure factors of the Si84 system measured at n = 113μm$^{-3}$ and increasing concentrations of NaOH. Curves are shifted for clarity. From top to bottom: $c_{NaOH}$ = 0 mmol/l, 0.22 mmol/l, 1.06 mmol/l, 2.28 mmol/l and 3.80 mmol/l. At the lowest and largest NaOH concentration the sample shows a fluid structure while at medium concentrations a bcc polycrystalline structure is appearing as indicated by the Miller indices.*

The detector resolution was 2048×2048 (with a pixel size of *79.1μm*). The detector integration time was chosen to be 10sec. For radially averaged Intensity $I(q)$ a conventional background and transmission correction was carried out before the structure factor $S(q)$ was calculated via $I(q)=I_0P(q)S(q)$, were $I_0P(q)$ is accessible at very large sodium hydroxide concentrations, $c_{NaOH}$, were particle interactions are sufficiently suppressed ($S(q)=1$).[31]

Extracted USAXS static structure factors are shown in Fig. 1 for constant $n = 113 μm^{-3}$ and various $c_{NaOH}$. At low and large $c_{NaOH}$ the sample shows fluid structure, while at medium concentrations it forms a bcc polycrystalline structure as indicated by the miller indices. A fcc structure was not identified but is expected at higher particle concentration. The phase transition is clearly recognized by the appearance of crystalline peaks. It was observed close to the phase boundaries the crystalline peaks become narrower and sharply indicating an increasing crystallite size (increasing long range order). Production of increasingly larger crystals can be understood in terms of an Avrami model of nucleation and growth, were close to the freezing at lower metastability the system changes to more growth influenced solidification process.[32]

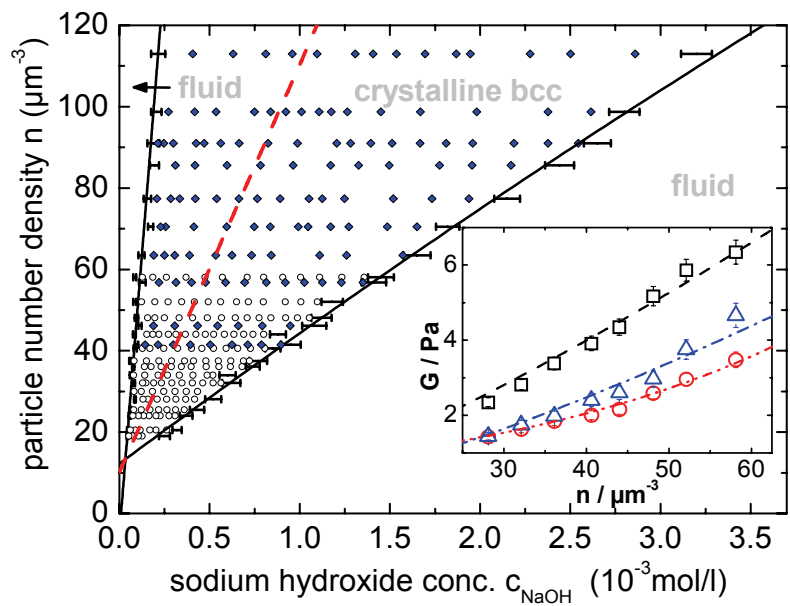

Fig. 2: (Color online) *Full phase diagram of charged silica (Si84) spheres in terms of the particle number density n (determined from the positions of the first peak of the structure factor) and the amount of added NaOH (determined from conductivity at known n.) Open symbols denote microscopy and light scattering data, closed symbols denote USAXS data. The*

*coexistence regions are indicated by horizontal bars. The system shows re-entrant melting due to charging (left boundary) and screening (right boundary). The central dashed line denotes the location of maximum interaction (largest shear moduli) which roughly coincides with the equivalence points of titration. The inset shows the shear moduli measured at different densities at the left phase boundary (up triangles) under condition of maximum interaction (squares) and at the right phase boundary (circles). The lines are fits to the data yielding $Z_{eff}$=253±15 at the left phase boundary, $Z_{eff}$=340±20 at maximum interaction and $Z_{eff}$=319±15 at the right phase boundary.*

The full phase diagram is shown in Fig. 2. Upon increasing the concentration of *NaOH* the particles charge up *via* the reaction $Si\text{-}OH + NaOH \rightarrow Si\text{-}O^- + Na^+ + H_2O$. Bare charge *Z*, conductivity charge *Z\** and elasticity charge $Z_{eff}$ increase. For sufficiently strong interaction the system freezes, with the vertical bars denoting the coexistence region. The dashed line in the centre of the crystalline region corresponds to the maximum effective charge $Z_{eff}$ and coincides with minimum conductivity. Beyond the equivalence point the concentration of *NaOH* increases. At low pH the silica dissolves to form monosilicate at concentrations of about *1%w/w* of the silica particles. Neglecting the small contribution of the monosilicate´s to the conductivity the excess concentration of *NaOH* is directly determined using Hessinger´s model.[16] With increasing $c_{NaOH}$ the interactions become screened at only slightly decreasing charge and the system melts again. This effect differs from that observed by Yamanaka et al.[22], who studied their sample only in a limited range of *6<pH<8* with little excess *NaOH* and attributed their re-entrant behaviour to the self screening effect upon charging. Here we have additional screening by the added electrolyte and hence may investigate the dependence of the phase behaviour for all three relevant interaction parameters.

To compare with the phase behaviour predicted by computer simulation we determined $Z_{eff}$ along both phase boundaries and at maximum interaction from the shear moduli. The inset in Fig. 2 shows the results for increasing *n*. With the best fit applied to the respective data sets with a fitting procedure described in details elsewhere[33], the mean effective charges were determined to be $Z_{eff}$=253±15 at the left phase boundary, $Z_{eff}$=319±15 at the right phase boundary and at maximum interaction we yield $Z_{eff}$=340±20. At this state the bare surface charge density of the *Si84* system is $\sigma = 3.4\mu C/cm^{-2}$ which is much smaller than for previously used particles and corresponds to *Z = 4500e*. It should be noted that elasticity charges are considerably smaller than corresponding conductivity charges $Z_\sigma$ indicating the massive presence of macro-ion shielding.[33]

Using these values we calculate the effective temperature $T^* = k_B T / V(\bar{d})$ and the coupling parameter $\lambda = \kappa \bar{d}$ via the screened Debye Hückel Potential $V(\bar{d})/k_B T = (Z_{eff}^2 \lambda_B) (\exp(\kappa a) / 1+\kappa a)^2 (\exp(-\kappa \bar{d}) / \bar{d})$ with the Bjerrum length $\lambda_B = 0.72 nm$, $\bar{d} = n^{-1/3}$ and the screening parameter $\kappa = \sqrt{4\pi k_B T \lambda_B \left(n Z_{eff} + 2 \times 10^3 N_A c_{NaOH}\right)}$, where the first term is the counter-ion concentration and the second accounts for the excess electrolyte. Ionic species stemming from the self dissociation of water and from the monosilicate formation are neglected. This allows to construct the state lines for our sample in the effective temperature – coupling plane introduced by RKG[18] (c.f. Fig. 3). We consider an increased $n$ at maximum interaction (circles) and the charging and screening processes at different constant $n$ and increasing *NaOH* concentration. The large symbols denote melting. In all three cases a good agreement with the theoretical expectation is observed. Thus one main result of the present study is the observation, that for charge variable particles the location of the melting transition can be precisely carried out, if effective charges which account for both counter-ion condensation and macro-ion shielding are used. In other words, many body effects strongly influence the phase behaviour of charged colloidal spheres.

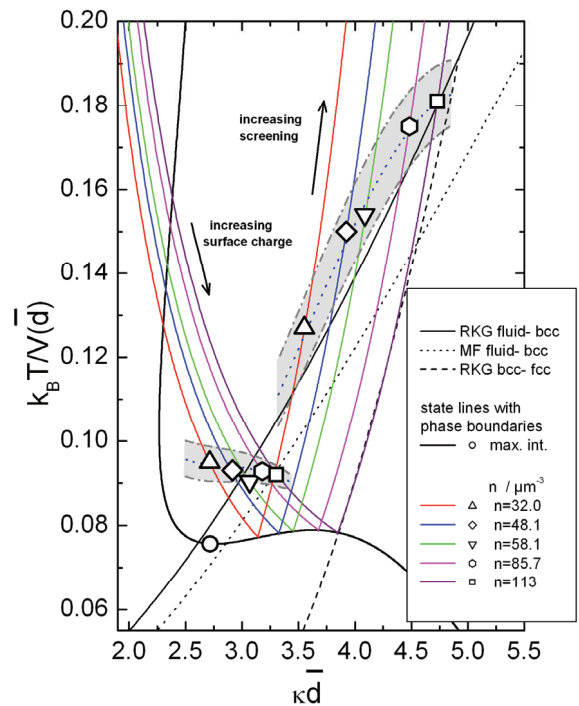

Fig. 3: *(Color online) Phase behaviour of the charge variable silica spheres in the effective temperature – coupling strength plane as introduced by RKG [18]. The solid line denotes the*

*universal melting line of RKG, the dotted line the later result of MF [19], and the dashed line gives the bcc/fcc transition. The large symbols denote the experimentally observed melting transition(s) at maximum interaction with increasing n (circle) and at constant n = 32.0μm$^{-3}$ (up triangles), 48.1μm$^{-3}$ (diamonds), 58.1μm$^{-3}$ (down triangles), 85.7μm$^{-3}$ (hexagons), 113μm$^{-3}$ (squares) with increasing NaOH concentration. The melting points at both phase boundaries are surrounded by error regions (light grey areas) accounting for the uncertainties in $Z_{eff}$, $c_{NaOH}$ and n. The thin coloured lines represent the corresponding state lines. At maximum interaction with increasing n the corresponding state line was calculated using the maximum effective charge $Z_{eff}$=340. The other state lines at constant n and increasing NaOH correspond to increasing surface charge (left state lines) and screening (right state lines).*

A few remarks are at place. First, the maximum interaction state line does not show a pronounced maximum, rather a saddle is observed and the melting line is crossed only once. This indicates graphically, that self screening is much smaller than in previous studies. Our re-entrant behaviour is due to excess electrolyte. Second, the data at the left phase boundary are close to both the original prediction of RKG and later results of MF.[19] Here within error bar we cannot discriminate between these alternatives. Whereas the right phase boundary transition better coincides with the RKG prediction. Third, however, for small $c_{NaOH}$ the melting points run nearly perpendicular to the fluid solid phase boundary and at large $c_{NaOH}$ the data lie systematically short above the prediction. Both are attributed to the neglect of the contributions of monosilicate. Fourth the largest concentration state line just touches the bcc/fcc melting transition, but evidence of this transition was neither gained from elasticity nor from scattering experiments. Possibly this transition is encountered at larger *n*. Finally, we here used an approach different to that of some theoretical work[25,26] who corrected the universal melting line downward. Instead we used a reduced effective charge yielding elevated state lines and could reproduce the prediction of RKG.

We have shown that re-entrant melting can be observed in charge variable colloidal spheres due to first charging then screening of interactions occurring upon increasing the amount of added *NaOH*. We demonstrated that the observed transitions could be well described with a screened Coulomb repulsion, if not only counter-ion condensation (as predicted by charge renormalization theory [8,10,15]) but furthermore also many body effects by macro-ion shielding are accounted for, which are accessible *via* the orientationally averaged elasticity. For our case we do not need to introduce like charge attractions discussed by other authors, as we observe a convergence between both effective charges neatly explained by a reduced repulsion range. We anticipate that in the long run a better modelling of the chemical processes occurring could even lead to a discrimination between different theoretical predictions based on purely repulsive interactions.

The authors like to thank J. Horbach for stimulating discussions and A. Timmann and S. Klein for experimental support. Financial support of the DFG (SPP1120, SPP1296, He1601/17, 23, 24 and Pa459/12, 13, 14) the MWFZ, Mainz and the DLR, Cologne is gratefully acknowledged.